\documentclass[journal,10pt,final]{IEEEtran}

\usepackage[T1]{fontenc}
\usepackage{cite}
\usepackage{url}
\usepackage{amsmath}
\usepackage{amssymb}
\usepackage[cmintegrals]{newtxmath}
\usepackage{subcaption}
\usepackage{dblfloatfix}
\usepackage{graphicx}
\usepackage{epstopdf}
\usepackage{xcolor}
\usepackage{mathtools}
\usepackage[detect-weight=true, binary-units=true]{siunitx}
\usepackage[inline]{enumitem}
\usepackage{lettrine}
\usepackage[boxruled,norelsize]{algorithm2e}

\makeatletter
\newcommand{\removelatexerror}{\let\@latex@error\@gobble}
\makeatother

\begin{document}
\title{A Profit-Maximizing Strategy of Network Resource Management for 5G Tenant Slices}

\author{Bin~Han,~\IEEEmembership{Member,~IEEE,}
	Di~Feng,
	Lianghai~Ji,~\IEEEmembership{Student Member,~IEEE,}
	and~Hans~D.~Schotten,~\IEEEmembership{Member,~IEEE}
	\thanks{\textit{B. Han and H. D. Schotten are with Institute for Wireless Communication and Navigation, Department of Electrical and Computer Engineering, University of Kaiserslautern, 67663 Kaiserslautern, Germany.}}
	\thanks{\textit{Emails: \{binhan,schotten\}@eit.uni-kl.de}}%
	\thanks{\textit{D. Feng is with Universitat Aut\`onoma de Barcelona and Barcelona GSE, Cerdanyola del Vall\`es, 08193 Spain.}}
	\thanks{\textit{Email: Di.Feng@e-campus.uab.cat}}%
}


%



\maketitle

\begin{abstract}
Supported by the emerging technologies of Network Function Virtualization (NFV) and network slicing, 5G networks allow tenants to rent resources from mobile network operators (MNOs) in order to provide services without possessing an own network infrastructure. The MNOs are therefore facing the problem of deciding if to accept or decline the resource renting requests they receive. This paper builds a stochastic model that describes the MNO's revenue and opportunity cost of accepting a contract, and therewith proposes a strategy that is analytically derived to maximize the expected profit at every decision.
\end{abstract}


%
\IEEEpeerreviewmaketitle

\section{Introduction}
\IEEEPARstart{N}{etwork} slicing was proposed by the \textit{Next Generation Mobile Networks (NGMN) Alliance} \cite{alliance20155g}, since then it has become one of the hottest topics in the filed of future 5\textsuperscript{th} Generation (5G) mobile communication networks. Generally, the concept of network slicing can be understood as creating and maintaining multiple independent logical networks (slices) on a common physical infrastructure, each slice operates a separate business service. Enabled and supported by the emerging technologies of software defined networks (SDN) and network function virtualization (NFV), network slicing exhibits great potentials, not only in supporting specialized applications with extreme performance requirements, but also in benefiting the mobile network operators (MNOs) with increased revenue \cite{rost2017network}.

As pointed out by \textit{Rost} et al. \cite{rost2017network}, a sliced mobile network manages its infrastructure and virtual resources in independent scalable slices, each slice runs a homogeneous service with simple business model. Thus, an MNO can dynamically and flexibly create, terminate and scale its slices to optimize the resource utilization for a better revenue or profit.

In a previous paper \cite{han2017modeling}, we have proposed a profit optimization model for sliced mobile networks that run in the traditional business mode: the MNOs with network resources implement the slices and provide all network services directly to their end-users. In this case, an MNO has full a priori knowledge about the service demand and the cost/revenue models of its every slice. It is able to scale the slices according to their profiting efficiencies, in order to achieve the maximal overall profit under the resource constraint. This is a classic multi-objective optimization problem (MOOP), in which the main challenge is to solve the optimum, or at least to find a satisfactory solution, with an affordable computing effort.

Unfortunately, this model does not apply to the slices run by tenants such as mobile virtual network operators (MVNOs), which are considered to play an important role in 5G networks \cite{rost2016mobile}. Tenants are third-parties that provide services without owning any network infrastructure, e.g. utility/automotive companies and over-the-top service providers such as \textit{YouTube}. To implement services, they have to be granted by MNOs with network resources, including radio / infrastructure resources and virtualized resource blocks. In legacy networks, every tenant makes its contractual agreement with the MNO(s), to pay a fixed and coarsely estimated annual/monthly fee for these resource sharing concepts. In the context of network slicing, in contrast, the resources are first bundled into slices before granted to tenants upon demand. Depending on the type, size and lifetime of granted slice, the fee is specified. This approach improves the sharing efficiency and the resource utilization rate. However, as such slices are maintained by tenants, the MNO has neither insight into their efficiencies of making revenue at end-users, nor authentication to rescale or terminate them during their lifetime. Instead, the MNO formulates the fee rate for different resource bundles, and chooses if to accept or decline the slice requests from tenants, like discussed in \cite{bega2017optimising}. In this case, the MNO cannot jointly optimize all slices in a fully dynamic approach, but only attempt to make the best decision for every received request, which is a problem of decision theory and operations research.

In this paper, we will focus on the case of tenant slices, and propose an economic model that evaluates the profit of an MNO to accept a certain slice request from tenant. Based on this model we propose a decision strategy to maximize the expected overall network profit at every decision step. The rest part of the paper is organized as follows: In Sec. \ref{sec:simp_model} we simplify the problem described above to an approximate business model. Then we build the profit model in Sec. \ref{sec:profit_model}, starting with an ideal simple case  and then approach to the complex reality step-by-step, on every step we deduce a profit-maximizing decision strategy from the proposed profit model. 
At the end we close the paper with our conclusion and outlooks in Sec. \ref{sec:conclusion}.

\section{Simplified Business Problem}\label{sec:simp_model}
\subsection{Fundamental Assumptions}\label{subsec:assumptions}
Our study begins with some basic assumptions and approximations on the business case. First of all, in most countries and regions, the mobile network infrastructure is controlled by only a few or even one single MNO, i.e. the network resource market is never perfectly competitive but highly oligarchy or monopoly. Hence, in this work we consider the case with only one MNO, ignoring the competition between different MNOs.

Second, the MNO holds a resource pool, which contains resources of certain types. Self-evidently, resource of every type is limited in amount. To rent these resources to tenants, a list of available contracts is provided by the MNO, every contract defined by a resource bundle, a contract period and a periodical payment. Every resource bundle is specified for a reference slice of certain type and size. We assume that the list of available contracts are predefined and remains consistent.

When a tenant requires resources to implement a slice, it selects one from the available contracts, requests to possess the corresponding resource bundle for the contract period. The MNO then decides if to accept or decline the request. Upon acceptance, the contract is confirmed and the tenant periodically pays the quota defined in the contract. If denied, the requested resource bundle will not be dedicated, and can be flexibly exploited for the MNO's own slices to make revenue. 

We consider that a confirmed contract cannot be terminated or modified within its period. We neglect the priority of contract renewals over new contract establishments, i.e. a tenant obtains no advantage for its future requests from the current contract. We also consider only nonelastic slices and neglect resource multiplex over slices, i.e. no resource can be allocated to multiple contracts simultaneously. Therefore, when accepting the current request, the MNO also loses some opportunity of accepting potential better deals in future.

The requests arrive stochastically. Usually, the arriving rate remains on a certain level and the intervals between different arrivals are independent from each other. Hence, it is reasonable to consider the number of arriving requests in a certain period as Poisson distributed. To simplify the model we consider an enough short unit period so that the request arrivals can be approximated as a Bernoulli process. 
We also assume that the MNO possesses full a priori knowledge about the statistics of arriving requests (resource bundle and contract period), which we consider as consistent.

\subsection{Model Setup}
To normatively describe the simplified business model above, we define the following sets, variables and mappings:

\begin{itemize}
	\item $\Psi= [0,1]^N$: a general $N$-dimensional non-negative Euclidean space to measure normalized network resource bundles in reference to the maximal resource pool, where $N$ is the number of resource types.
	\item $\psi_t\in\Psi$: the normalized measure of idle resource pool available at discrete time $t\in\mathbb{N}$.
	\item $\Omega_t\subset\Psi$: the finite, discrete set of resource bundles defined by all contract options provided by the MNO with its idle resources at time $t$.	Each element in $\Omega_t$ is a possible resource bundle requested at time $t$. The case of no request arrival is considered as a null-request, i.e. $\omega_\textrm{null}=0^N\in\Omega_t,\forall t\in\mathbb{N}$. We also consider that the entire resource pool is idle at $t=0$ so that $\Omega_0$ is a predefined set of resource bundles defined by the list of all available contracts, and generally we have a generation function:
	\begin{equation}
		\Omega_t=G(\psi_t)=\{\omega|\omega\le\psi_t\}\cap\Omega_0, \forall t\in\mathbb{N}
		\footnote{Note that here with $\omega\le\psi_t$ we denote $\omega$ is not greater than $\psi_t$ in any of the $N$ dimensions.}
	\end{equation}
	\item $\mathbb{T} \subset \mathbb{N}^+$: the finite, discrete set of contract periods in all contract options, $\inf(\mathbb{T})=1$ denotes a unit time period.
	\item $P \subset \mathbb{R}$: the finite, discrete set of payments defined by all contract options.
	\item $\Omega_0 \xrightarrow{g} [0,1]$: a probability measure on $\Omega_0$ representing the probabilities of upcoming request for different contracts in each unit time period.
	\item $\Omega_0\times\mathcal{P}(\Omega_0) \xrightarrow{f} [0,1]$: a probability measure representing the probabilities of upcoming request for different contracts that can be supported by a given idle resource pool\footnote{$\mathcal{P}(\Omega_0)$ denotes the power set of $\Omega_0$}. We consider requests for oversize contracts beyond the support of current idle resource pool as equivalent of null contracts, hence generally:
		\begin{equation}
		\begin{split}
			f(\omega,\Omega)
			=\begin{cases}
			0&\omega>\sup(\Omega)\\
			g(\omega_{\textrm{null}})+\sum\limits_{\mu>\sup(\Omega)}g(\mu)&\omega=\omega_{\textrm{null}}\\
			g(\omega)&\textrm{otherwise}
			\end{cases}
		\end{split}
		\end{equation}
	\item $\Psi\times\Omega_0 \xrightarrow{F} \Psi$: the decision strategy that determines if to accept the arrived requests. It maps from the current idel resource pool and a received request to the idle resource pool at the next period, i.e. $F(\psi_t,\omega_t)=\psi_{t+1}$.
	\item $\Omega_0\times \mathbb{T} \xrightarrow{p} P$: a pricing function that maps the resource bundle and contract period $(\omega,T)$ to the corresponding periodical payment, the payment of a null contract is zero i.e. $p(\omega_\textrm{null},T)=0,\forall T\in\mathbb{T}$.
	\item $\Omega_0\xrightarrow{q}\mathbb{R}$: a function that maps a resource bundle to the corresponding revenue it can generate within one unit time period through its deployment in MNO's own slices. In this work we consider $q$ as a linear function of $\omega$.
\end{itemize}


\section{Profit Model and Decision Strategy}\label{sec:profit_model}
Now we analyze the MNO's profit of accepting a resource request from tenant. In the following part of the paper, we use the term $(\psi_t,\omega_t,T_t)$ to represent the request arriving at $t$.

\subsection{Two-Step Decision, Non-Expiring Contract}
We start with a simple two-step model, where both the MNO and the tenant only act two unit periods, i.e. $t\in\{0,1\}$. We also assume that no contract will expire, which means $T_t=2-t$.
Now for the two periods we have two idle resource pools $\psi_0,\psi_1$. As there is no contract expiry, no resource is released at $t=1$ so that
\begin{align}
	\psi_1=\begin{cases}
		\psi_0-\omega_0&(\psi_0,\omega_0,2)\textrm{ accepted};\\
		\psi_0&\textrm{otherwise}.
	\end{cases}
\end{align}
Obviously, as the MNO only operates two periods, it should accept any contract request at $t=1$ as long as its resource pool supports. So the focus is on the decision at $t=0$. Given any contract request $(\psi_0,\omega_0,2)$ arrived at $t=0$, the expected payoff of accepting this request is
\begin{equation}
	\Gamma_1(\psi_0,\omega_0)=(1+\beta)p(\omega_0,2)-C_1(\psi_0,\omega_0),
\end{equation}
where $\beta\in(0,1)$ is the discount factor to describe the time value of money (TVM) \cite{needles2013managerial}, which is determined by the capital market. $C_1$ is the \textit{Opportunity Cost} (OC) of this contract:
\begin{equation}
\begin{split}
&C_1(\psi_0,\omega_0)=(1+\beta)q(\omega_0)\\
&+\beta\sum\limits_{\omega\in\Omega_0}[f(\omega,\Omega_0)-f(\omega,G(\psi_0-\omega_0)]p(\omega,1).
\end{split}
\end{equation}
Obviously, to maximize the profit, the MNO is supposed to follow the optimal strategy of binary decision:
\begin{equation}
		\psi_{1}=F_{1,\textrm{opt}}(\psi_0,\omega_0)=
		\begin{cases}
			\psi_0-\omega_0&\Gamma_1(\psi_0,\omega_0)\ge0;\\
			\psi_0&\textrm{otherwise},
		\end{cases}
\end{equation}
where case 1 denotes acceptance and case 2 for declination.

\subsection{Multi-Step Decision, Non-Expiring Contract}
Then we progress towards the multi-step model, where the MNO and the tenant act $t_{\max{}}$ unit periods. Once again, as of this step we still consider non-expiring contracts, so that the request arriving at $t$ has a contract period of $T_t=t_{\max{}}-t$ and hence a periodical payment of $p(\omega_t,t_{\max{}}-t)$. Taking 
it into account that time present value of any future payment $x$ in $\Delta t$ periods from current under a discount factor $\beta$ is $x\beta^{\Delta t}$,
we can compute the total present value of all payments for the requested contract $(\psi_t,\omega_t,t_{\max{}}-t)$ as
\begin{equation}
	p_{t_{\max{}}}(\omega_t)=\sum_{\tau=t}^{t_{\max{}}-1} \beta^{\tau-t}p(\omega_t,t_{\max{}}-t).
	\label{equ:multi-step_non-expiring_payments}
\end{equation} 
Similarly, the present value of exploiting the resource bundle $\omega_t$ on the MNO's own slice for $t_{\max{}}-t$ can be computed as
\begin{equation}
	q_{t_{\max{}}}(\omega_t)=\sum\limits_{\tau=t}^{t_{\max{}}-1}\beta^{\tau-t}q(\omega_t).
	\label{equ:multi-step_non-expiring_lost_revenue}
\end{equation}
Thus, the OC of any request arriving at $t$ is
\begin{equation}
\begin{split}
	&C_{t_{\max{}}}(\psi_t,\omega_t)=q_{t_{\max{}}}(\omega_t)+\sum_{\tau=t}^{t_{\max{}}-1} \beta^{\tau-t+1}
	\\&\times\sum\limits_{\omega\in G(\psi_\tau)}[f(\omega,\Omega_\tau)-f(\omega,G(\psi_\tau-\omega_t))]p(\omega,t_{\max{}}-t).
\end{split}
\label{equ:multi-step_non-expiring_oc}
\end{equation}
Therefore, let
\begin{equation}
	\Gamma_{t_{\max{}}}(\psi_t,\omega_t)=	p_{t_{\max{}}}(\omega_t)-	C_{t_{\max{}}}(\psi_t,\omega_t),
	\label{equ:multi-step_non-expiring_payoff}
\end{equation}
the profit-maximizing decision by the MNO should be
\begin{equation}
		\psi_{t+1}=F_{t_{\max{}},\textrm{opt}}(\psi_t,\omega_t)=
		\begin{cases}
			\psi_t-\omega_t&\Gamma_{t_{\max{}}}(\psi_t,\omega_t)\ge0;\\
			\psi_t&\textrm{otherwise}.
		\end{cases}
		\label{equ:multi-step_non-expiring_decision}
\end{equation}
Note that as the OC is not promised to be convex 
about the decision sequence $[\psi_{0},\dots,\psi_{t_{\max{}}-1}]$, (\ref{equ:multi-step_non-expiring_decision}) is not guaranteed to achieve the global optimum of $C_{t_{\max{}}}(\psi_t,\omega_t)$, but a local maximum. 
Besides, according to (\ref{equ:multi-step_non-expiring_oc}),  the OC $C_{t_{\max{}}}$ depends on the expected idle resource pool in future $\psi_{\tau>t}$, which is determined by the target decision strategy $F_{t_{\max{}},\textrm{opt}}$, whose solution (\ref{equ:multi-step_non-expiring_decision}) relies on $C_{t_{\max{}}}$. This closed loop encourages to apply an iterative approach, as briefly described in Fig. \ref{fig:iterative_solution}.
\begin{figure}[!h]
	\removelatexerror
	\begin{algorithm}[H]
		\footnotesize
		Start with initial $\Omega_0,\psi_0, g(\omega)$, a default strategy $F_{t_{\max{}}}^0$, a maximal number of iterations $i_{\max{}}$ and aconvergence threshold $\gamma$. \;
		\For(){$i = 0$ to $i_{\max{}}$}
		{
			Generate a request sequence according to $g(\omega)$\;
			Determine the decision sequence according to $F_{t_{\max{}}}^i$\;
			Compute the OC $C_{t_{\max}}^i(\psi_0,\omega)$ under the new decision sequence for all $\omega\in\Omega_0$ according to (\ref{equ:multi-step_non-expiring_oc})\;
			Update the strategy $F_{t_{\max{}},\textrm{opt}}^i$ with the OC according to (\ref{equ:multi-step_non-expiring_decision})\;
			\uIf{$i\ge1,\sum\limits_{\omega\in\Omega_0}{|C_{t_{\max{}}}^{i+1}(\psi_0,\omega)-C_{t_{\max{}}}^{i}(\psi_0,\omega)|}<\gamma$}{
				\Return $F_{t_{\max{}},\textrm{opt}}^{i}$\;
			}
			\Else{
				Update the strategy: $F_{t_{\max{}}}^{i+1}\gets F_{t_{\max{}},\textrm{opt}}^i$\;
			}
		}
		\Return $F_{t_{\max{}},\textrm{opt}}^{i_{\max}}$\;
	\end{algorithm}
	\caption{An iterative algorithm to approximate the target strategy. In each iteration the strategy is tested and correspondingly updated with a simulated request sequence, and the convergence is evaluated by the estimated total OC of the simulated request sequence under the updated strategy.}
	\label{fig:iterative_solution}
\end{figure}

As $C_{t_{\max}}$ has non-negative terms and bounded partial sums, it converges with increasing $t_{\max}$. Therefore the sequence $(C_{t_{\max{}}}^i)_{i\in\mathbb{I}}$ is bounded where $\mathbb{I}=\{1,2,\dots,i_{\max{}}\}$. Thus, according to the Bolzano-Weierstrass theorem \cite{bartle2000introduction} it always has a converging subsequence $(C_{t_{\max{}}}^j)_{j\in\mathbb{J}\subseteq\mathbb{I}}$. By selecting a reasonable $i_{\max{}}$ we can renew $\mathbb{I}=\mathbb{J}$ in order to construct a converging sequence $(C_{t_{\max}}^i)$, so that the iterative algorithm converges to a limit as $i$ approaches to $i_{\max{}}$.\footnote{Herewith we have analytically derived the convergence, but the converging speed must be numerically evaluated, which shall be a follow-up work.}

In real world, the MNO and tenants shall be considered as long-term or even eternally operating, i.e. $t_{\max{}}\to+\infty$. According to (\ref{equ:multi-step_non-expiring_decision}), the sequence $(\psi_0, \psi_1,\dots)$ monotonically decreases, and $G(\psi_t)$ converges to an empty set:
\begin{equation}
	\lim\limits_{t\to+\infty}G(\psi_t)=\emptyset.
	\label{equ:converging_G}
\end{equation}
Therefore we know that $C_{t_{\max{}}}(\psi_t,\omega_t)$ is bounded as $t_{\max{}}\to+\infty$. Meanwhile, as $p$ is bounded and $\beta\in(0,1)$, $p_{t_{\max{}}}$ also converges to a bounded value as $t_{\max{}}\to+\infty$, so the proposed approach also applies to the infinite-step case\footnote{Nevertheless, as the MNO is usually interested in a short-term or intermediate-term (e.g. monthly or annual) profit, instead of the long-term overall profit till forever, it is still practical to artificially set a finite $t_{\max{}}^*$.\label{footnote:artificial_tmax}}.

Generally, in the case of non-expiring contract, at every step of decision, the impacts of all historical decisions about previous requests are completely reflected in the current idle resource pool without any extra influence in the future. Hence, the non-expiring contract model is a Markov model, and the proposed decision strategy is also Markovian as well.
%
%
%
%

\subsection{Multi-Step Decision, Expiring Contract}
Subsequently we bring the issue of contract expiry into our discussion. Consider flexible contract periods $T_t<t_{\max{}}-t$ permitted, the resource bundle assigned to it will be released to the MNO's resource pool after the expiry. Thus, the idle resource pool size $\psi_t$ is not monotonically decreasing with $t$, but jointly determined by all previous requests and decisions during $[0,t-1]$. Hence, the model becomes non-Markovian and the Markovian decision strategy (\ref{equ:multi-step_non-expiring_decision}) does not apply as it lacks information about contract periods and previous decisions.

As a solution to this, given an arbitrary request $(\psi_\tau,\omega_\tau,T_\tau)$ arriving at $\tau$, we denote the indicator function
\begin{align}
	I_\tau(t,T_\tau)=\begin{cases}
		1&\tau\le t\le \tau+T_\tau, (\psi_\tau,\omega_\tau,T_\tau)\textrm{ accepted};\\
		0&\textrm{otherwise}
	\end{cases}
\end{align}
to represent its validity at time $t$. Thus,  we can represent the resource bundle reserved for it at any time $t$ as
\begin{equation}
	\omega_t^{T_\tau}=\omega_tI_\tau(t,T_\tau)
\end{equation}
Then we define the scalar
\begin{equation}
\tilde{\omega}_t=\sum\limits_{\tau=0}^{t-1}\omega_t^{T_\tau}
\end{equation}
to track and aggregate the resource bundles currently reserved by all previously accepted contracts.
Thus, instead of $(\psi_t, \omega_t, T_t)$, now we use $(\tilde{\omega}_t, \omega_t, T_t)$ to represent a contract request. 
the equations (\ref{equ:multi-step_non-expiring_payments}) and (\ref{equ:multi-step_non-expiring_lost_revenue}) become
\begin{align}
	p_{t_{\max{}}}^{T_t}(\omega_t)&=\sum\limits_{\tau=t}^{t+T_t-1}\beta^{\tau-t}p(\omega_t,T_t),\\
	q_{t_{\max{}}}^{T_t}(\omega_t)&=\sum\limits_{\tau=t}^{t+T_t-1}\beta^{\tau-t}q(\omega_t),
\end{align}
respectively, and thus the OC of accepting the request is
\begin{align}
	\begin{split}
	&C_{t_{\max{}}}^{T_t}(\tilde{\omega}_t,\omega_t)=q_{t_{\max{}}}^{T_t}(\omega_t)+\sum_{\tau=t}^{t_{\max{}}-1} \beta^{\tau-t+1}
	\\&\times\sum\limits_{\omega\in G(\psi_\tau)}[f(\omega,\Omega_\tau)-f(\omega,G(\psi_0+\tilde{\omega}_\tau-\omega_t^{T_t}))]p(\omega,T_t).
	\end{split}
\end{align}
Defining the payoff function
\begin{equation}
		\Gamma^{T_t}_{t_{\max{}}}(\tilde{\omega}_t,\omega_t)=	p^{T_t}_{t_{\max{}}}(\omega_t)-C^{T_t}_{t_{\max{}}}(\tilde{\omega}_t,\omega_t),
\end{equation}
the non-Markovian profit-maximizing decision strategy is
\begin{equation}
	F_{t_{\max{}},\textrm{opt}}^{T_t}(\psi_t,\omega_t)=
		\begin{cases}
			\psi_0+\tilde{\omega}_{t+1}-\omega_t&\Gamma_{t_{\max{}}}^{T_t}(\psi_t,\omega_t)\ge0;\\
			\psi_0+\tilde{\omega}_{t+1}&\textrm{otherwise}.
		\end{cases}
\end{equation}

In finite-step cases where $t_{\max{}}$ is finite, the number of possible sequences $((\omega_t, T_t))_{0\le t\le t_{max{}}}$ is limited so that $C_{t_{\max{}}}^{T_t}$ is bounded and the iterative approach in Fig. \ref{fig:iterative_solution} still applies. However, when $t_{\max{}}\to+\infty$, as $(\psi_t)_{t\in\mathbb{N}}$ is not monotonic about $t$, (\ref{equ:converging_G}) fails to hold and no more convergence is guaranteed. In this case, an artificial finite $t_{\max{}}^*$ is needed, like we did in the footnote \ref{footnote:artificial_tmax}.

\section{Discussions}
So far we have closed our study on the MNO's decision strategy under the assumptions in Sec. \ref{subsec:assumptions}. Nevertheless, concerning the strength of our assumptions, some discussions about their feasibilities may be necessary.

The first concern can be raised by the monopoly model with only one MNO, because the tenants are may fail to obtain resources to maintain services in this case. Practically, a pre-ordering mechanism can be applied to allow early requests before the due of slice creation. Thus, if a request is declined, the tenant is still able to reattempt with another contract option with better chance of acceptance. Moreover, it is true that in practice there is  usually not only one but several MNOs, i.e. the market is actually oligarchy. While providing the tenant more alternative options when its request is rejected by one MNO, this fact does not necessarily conflict with our results under the monopoly assumption, because oligarchy markets differ from monopoly ones only in the supply, demand and pricing mechanism, but not in the decision making logic \cite{mas1995microeconomic}, which we focus on in this paper. Nevertheless, the competition between MNOs in oligarchy markets worths further study.

Another doubt may arise about the exclusion of unexpected contract termination and modification. Certainly they are ignored here for model simplification, and can be eventually involved in future work by introducing another random variable with corresponding statistics. Similarly, in this work we have ignored the preference in renewing old expiring contracts over making new ones. For a better approximation to reality we can consider all contracts as non-expiring, and apply the random termination event instead to describe the tenant cancellation.

At last, the assumption that MNO possesses full a priori knowledge about the request statistics may be argued. In practice, although the a priori model is hard to obtain, it can be estimated in a Bayesian approach from the a posteriori historical records that every MNO keeps, as long as it remains consistent. Even in case of non-stationarity, short-term consistence can still be approximated with periodical updates.

\section{Conclusion}\label{sec:conclusion}
In this work, in an operations research perspective we have investigated the 5G network resource management problem of creating tenant slices upon request, aiming at a strategy that maximizes the expected MNO revenue at every binary decision. Different cases have been studied and an iterative algorithm proposed. The convergence of proposed method has been mathematically derived. For future works, numerical experiments are expected to evaluate the converging speed and the revenue gain of the proposed algorithm in both short and long terms, elastic slices shall also be considered to enable resource multiplexing between slices .

\ifCLASSOPTIONcaptionsoff
  \newpage
\fi

\bibliographystyle{IEEEtran}
\bibliography{references}

%

%
%

%




\end{document}